\documentclass[12pt]{kluwer}    
\usepackage{epsf, rotate}

\begin{document}

\begin{article}
\begin{opening}

\title{Evolution of Isolated Neutron Stars}
\author{Sergei \surname{Popov}}
\runningauthor{Sergei Popov}
\runningtitle{Evolution of Isolated Neutron Stars}
\institute{Sternberg Astronomical Institute}
\date{November 10, 2000}

\begin{abstract}

 In this paper we briefly review our recent results on evolution 
and properties of isolated neutron stars (INSs) in the Galaxy.

As the first step we discuss stochastic period evolution of INSs.
We briefly discuss how an INS's spin period evolves under influence of interaction
with turbulized interstellar medium.

To investigate statistical properties of the INS population we
calculate a {\it census} of INSs in our Galaxy. 
We infer a lower bound for the mean  kick velocity of NSs,
$ <V> \sim $(200-300)  ${\rm km\,s^{-1}}$.
The same conclusion is reached for both a constant magnetic field
($B\sim 10^{12}$ G) and for a magnetic field decaying exponentially with a
timescale $\sim 10^9$ yr. These results,
moreover, constrain the fraction of low velocity NSs, which
could have escaped pulsar statistics, to $\sim$few percents.

Then we show that for exponential field decay 
the range of minimum  value of magnetic moment, $\mu_b$:
$\sim 10^{29.5}\ge \mu_b \ge 10^{28} \, {\rm G}\, {\rm cm}^3$, 
and the characteristic  decay time, $t_d$:
$\sim 10^8\ge  t_d \ge 10^7\, {\rm yrs}$, can be excluded
assuming the standard
initial magnetic momentum, $\mu_0=10^{30} \, {\rm G}\, {\rm cm}^3$,
if accreting INSs are observed.
For these parameters
an INS would never reach the stage of accretion from the interstellar
medium even for a low space velocity of the star
and high density of the ambient plasma. 
The range of excluded parameters increases for lower values of $\mu_0$.

It is shown that old accreting INSs  become more abundant than young cooling
INSs at X-ray fluxes below $\sim 10^{-13}$ erg cm$^{-2}$ s$^{-1}$.
We can predict that about one accreting INS per square
degree should be observed at the {\it Chandra} and {\it Newton} flux limits
of  $\sim 10^{-16}$ erg cm$^{-2}$ s$^{-1}.$
The weak {\it ROSAT} sources, associated with INSs,
can be young cooling objects, if the NSs birth rate
in the solar vicinity during the last $\sim 10^6$ yr was much higher than
inferred from radiopulsar observations.

\end{abstract}

\end{opening}

\section{Introduction}

Despite intensive observational campaigns, no irrefutable identification 
of an isolated accreting neutron star (NS) has been presented so far. 
Six soft sources have been found in ROSAT fields which are most 
probably associated to  isolated radioquiet NSs. 
Their present X-ray and optical data however do not allow an unambiguous
identification of the physical mechanism responsible for their emission. 
These sources can be powered either by accretion of the interstellar gas 
onto old ($\approx 10^{10}$ yr) NSs or by the release of internal energy 
in relatively young ($\approx 10^6$ yr) 
cooling NSs (see \cite{t2000} for a recent review). 
The ROSAT candidates, although relatively bright 
(up to $\approx 1 \ {\rm cts\,s}^{-1}$), are
intrinsically dim and their inferred luminosity 
($L\approx 10^{31} \ {\rm erg\, s}^{-1}$)
is near to that expected from either a close-by cooling NS or 
from an accreting NS among the most luminous.
Their X-ray spectrum
is soft and thermal, again as predicted for both accretors and coolers 
\cite{t2000}.
Up to now only two optical counterparts have been identified (RXJ 1856,
\cite{wm97}; RXJ 0720,  \cite{kk98}) and in both cases an
optical excess over the low-frequency tail of the black body X-ray spectrum 
has been reported.
While detailed multiwavelength observations 
with next-generation instruments may indeed be
the key for assessing the true nature of these sources, 
other, indirect, approaches may be
used to discriminate in favor of one of the two scenarios proposed so far.

Since early 90$^s$, when in \cite{tc91} it was suggested to search for IANSs
with ROSAT satellite, several investigations on INSs have been done (see
for example \cite{mb94}, \cite{mann96}, and \cite{t2000} for a review).
Here we present our recent results in that field.

\section*{Stochastic evolution of isolated neutron stars}

 One can distinguish four main stages of INS evolution: Ejector, Propeller,
Accretor and Georotator (see \cite{l92} for more detailes). Destiny of an
INS depends on its spin period, magnetic field, spatial velocity and
properties of the interstellar medium (ISM). Schematically typical
evolutionary paths can be shown on p-y diagram, where $y=\dot M/\mu^2$ --
gravimagnetic parameter (Fig.~1).

\begin{figure}[h!]
\epsfxsize=10truecm
\centerline{{\epsfbox{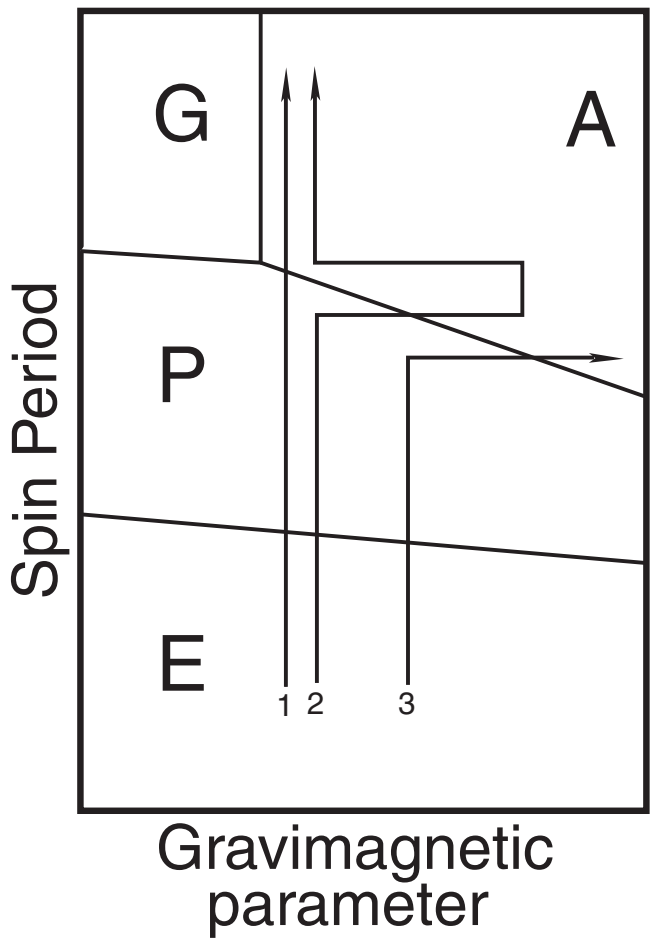}}}
\caption{P-y diagram. Three evolutionary tracks are shown:
1 -- evolution with constant field in constant medium.
The NS spins down passing through different stages (Ejector,
Propeller, Accretor); 2 -- passing through a giant molecular cloud.
The accretion rate is increased for some time, and the NS makes a loop
on the diagram;
3 -- evolution with field decay. The NS appears as Accretor
due to fast field decay. 
}
\end{figure}

 An INS normally is born as Ejector. Then it spins down, and appear as
Propeller. At last it can reach the stage of accretion,
if its spatial velocity is low enough, or if the magnetic field is high.

 As an INS evolves at the Accretor stage it more and more
feels the influence of the turbulized nature of the ISM (Fig.~2).
Initially isolated accreting NS spins down due to magnetic breaking,
and the influence of accreted angular momentum is small.
At last, at the stage of accretion, when the INS spun down enough,
its rotational evolution is mainly governed by the turbulence,
and the accreting angular moment fluctuates.

\begin{figure}[h!]
\epsfxsize=10truecm
\centerline{{\epsfbox{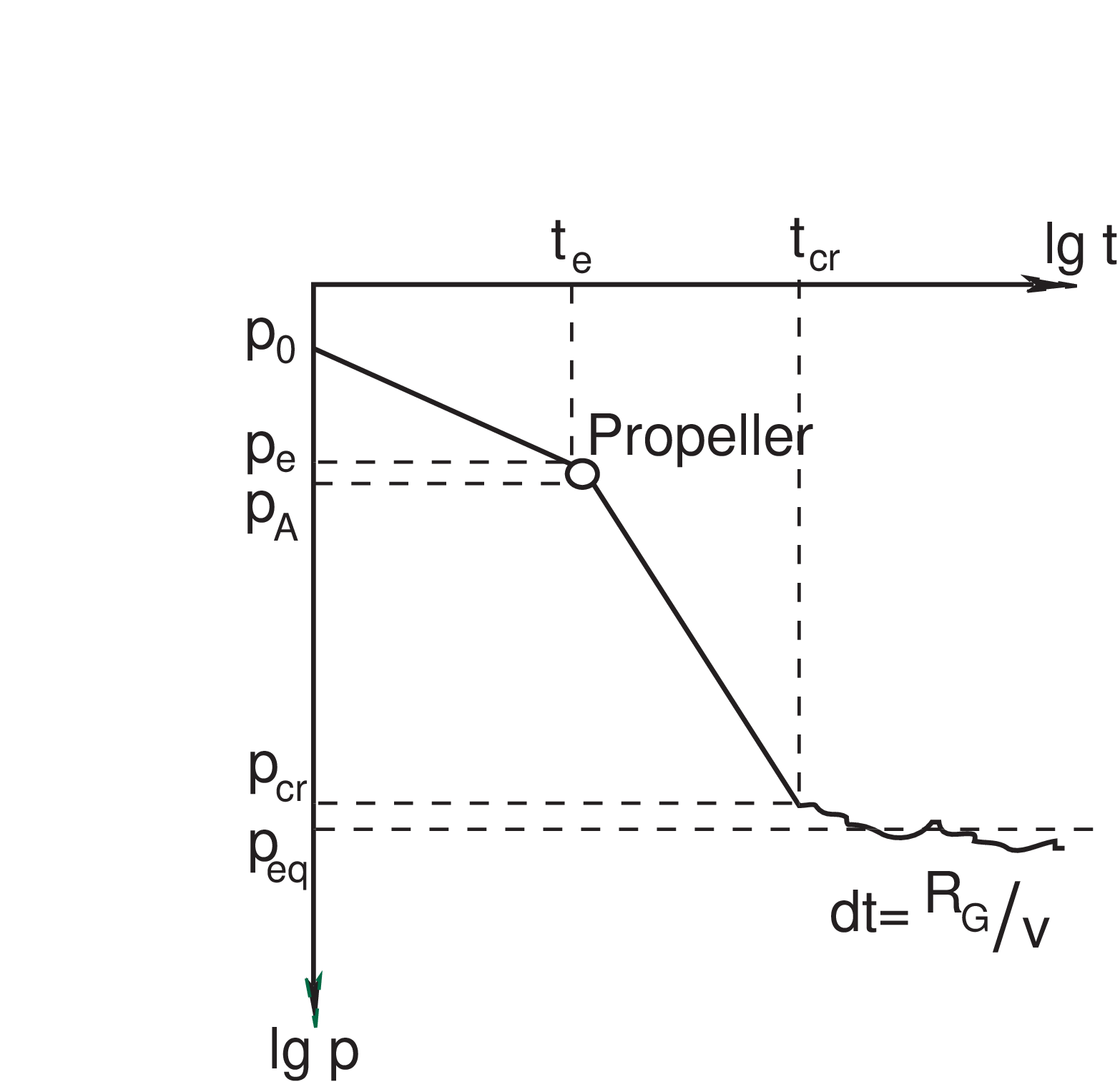}}}
\caption{Evolution of an isolated NS in turbulized interstellar medium.
After spin-down at the Ejector stage (spin-down up to $p=p_E$) and
short Propeller stage (shown with a
circle) the NS appears as accretor ($p>p_A$). Initialy magnetic breaking is more
important than accretion of the angular momentum from the interstellar
medium, and the NS constantly spins down. Then at $t=t_{cr}$
magnetic breaking and 
turbulent spin-up/spin-down become comparable, and spin period starts
to fluctuate coming to some average ("equilibrium") value, $p_{eq}$.
Typical timescale for fluctuations is $dt=R_G/v$, $R_G$ -- radius of
gravitational capture, $v$ -- spatial velocity. 
}
\end{figure}

 Observations of $p$ and $\dot p$ of isolated accreting NSs can provide us
not only with the information about INSs properties, but also about
different properties of the ISM itself.

\section*{Neutron Star Census}

We have investigated, \cite{p2000a}, how the present distribution of
NSs in the different stages (Ejector, Propeller,
Accretor and Georotator, see \cite{l92}) depends on the star mean velocity at
birth (see Fig.~3). The fraction of
Accretors  was used to estimate the number of sources within 140
pc from the Sun which should have been detected by ROSAT. Most
recent analysis of ROSAT data indicate that no more than $\sim 10$
non--optically identified sources can be accreting old INSs. This
implies that the 
average velocity of the INSs population at birth has to
exceed $\sim 200 \ {\rm km\, s^{-1}}$, a figure which is
consistent with those derived from radio pulsars statistics. We
have found that this lower limit on the mean kick velocity is
substantially the same either for a constant or a decaying
$B$--field, unless the decay timescale is shorter than $\sim 10^9$
yr. Since observable accretion--powered INSs are slow objects, our
results exclude also the possibility that the present velocity
distribution of NSs is richer in low--velocity objects with
respect to a Maxwellian. The paucity of accreting INSs seem
to lend further support in favor of NSs as
fast objects.

\begin{figure}[h!]
\epsfxsize=10cm
\centerline{\rotate[r]{\epsfbox{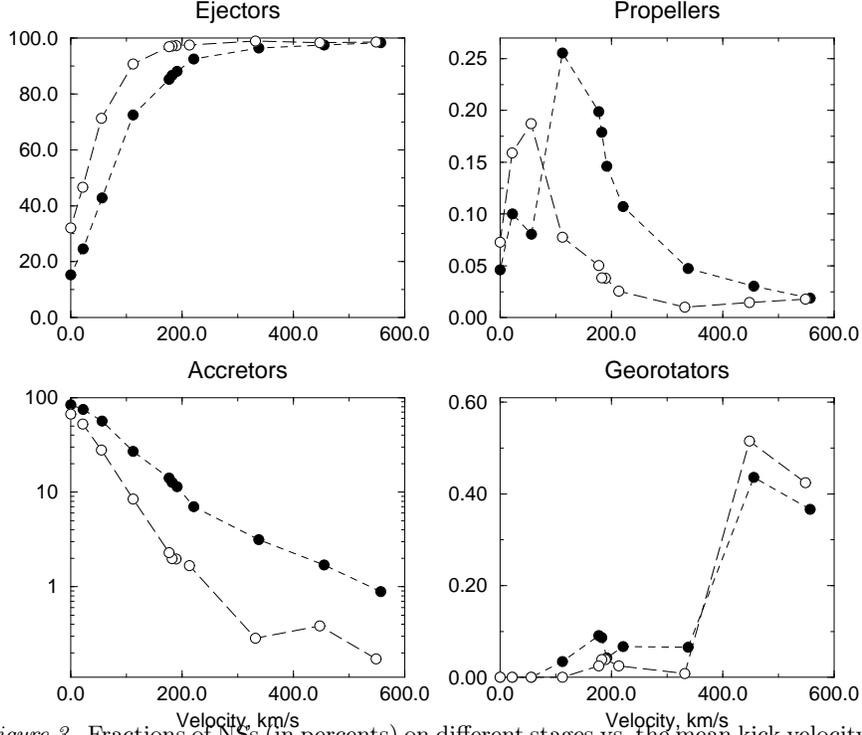}}}
\caption{Fractions of NSs (in percents) 
on different stages vs. the mean kick
velocity for two values of the magnetic moment: 
$\mu_{30}=0.5$ (open circles) and $\mu_{30}=1$
(filled circles). Typical statistical uncertainty for ejectors and
accretors is $\sim $ 1-2\%. Figures are plotted for constant magnetic field.}
\end{figure}

\section*{Magnetic Field Decay}

Magnetic field decay can operate in INSs. Probably, some
of observed ROSAT INS candidates represent such examples (\cite{kp97},
\cite{w97}) 
 We tried to evaluate the region of parameters which is
excluded for models of the exponential
magnetic field decay in INSs using
the possibility that some of  ROSAT 
soft X-ray sources are indeed old AINSs.  

In this section  we follow the article \cite{pp2000}.

\begin{figure}[h!]
\epsfxsize=10cm
\centerline{\rotate[r]{\epsfbox{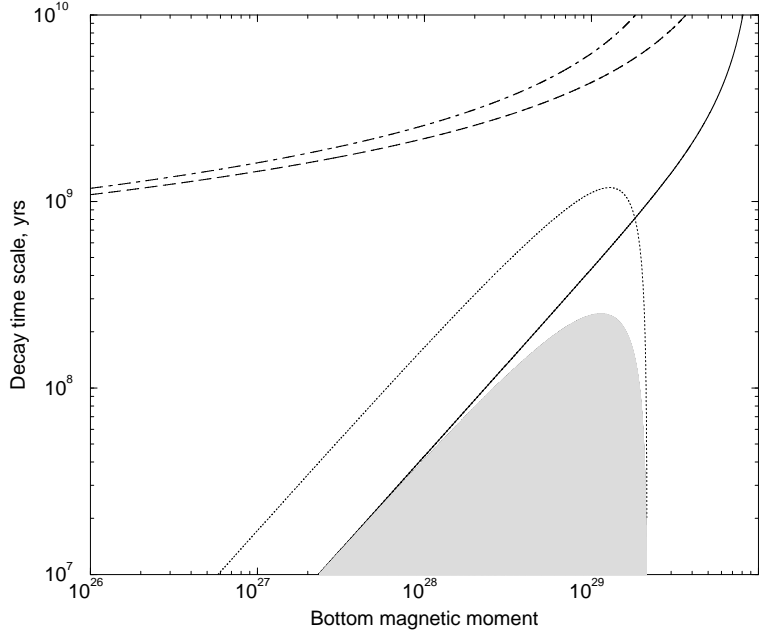}}}
\caption{ The characteristic time scale of the magnetic field decay, $t_d$, 
vs. bottom magnetic moment, $\mu_b$.
In the hatched region Ejector life time, $t_E$, 
is greater than $10^{10} {\rm yrs}$.
The dashed line corresponds to $t_H=t_d\cdot \ln \left( \mu_0/\mu_b
\right)$, where $t_H=10^{10}$ years. The solid line corresponds to
$p_E(\mu_b)=p(t=t_{cr})$, where $t_{cr}=t_d\cdot \ln \left(
\mu_0/\mu_b \right)$. Both the lines and hatched region
are plotted for $\mu_0=10^{30} {\rm G} \, {\rm cm}^{-3}$. 
The dash-dotted line is the same as the dashed one, 
but for $\mu_0=5\cdot 10^{29} 
\, {\rm G} \, {\rm cm}^3$.
The dotted line shows the border of the ``forbidden'' region for 
$\mu_0=5\cdot 10^{29}  \, {\rm G} \, {\rm cm}^3$. 
}
\end{figure} 

Here the field decay is assumed to have an exponential shape:

\begin{equation}
\mu=\mu_0\cdot e^{-t/t_d}, \, {\rm for} \, \mu > \mu_b
\label{eq:mu(t)}
\end{equation} 
where $\mu_0$ is the initial magnetic moment 
($\mu=\frac12 B_p R_{NS}^3$, here $B_p$ is the polar magnetic field,
$R_{NS}$ is the NS radius), $t_d$ is the characteristic time
scale of the decay, and $\mu_b$ is the bottom value of the 
magnetic momentum which is reached at the time 

\begin{equation}
t_{cr}=t_d\cdot \ln\left( \frac{\mu_0}{\mu_b} \right)
\end{equation}
and does not change after that. 

The intermediate values of $t_d$ ($\sim 10^7-10^8 \, {\rm yrs}$)
in combination with the intermediate values of
$\mu_b$ ($\sim 10^{28}-10^{29.5} \, {\rm G} \, {\rm cm}^3$)
for $\mu_0=10^{30} \, {\rm G}\, {\rm cm}^3$
can be excluded for progenitors of isolated accreting NSs
because NSs with such parameters would always remain
on the Ejector stage and never pass to the accretion stage
(see Fig.~4).
Even if all modern candidates are not accreting objects,
the possibility of limitations of magnetic field
decay models based on future observations of isolated accreting NSs
should be addressed. 

 For higher $\mu_0$ NSs should reach the stage of Propeller (i.e. $p=p_E$,
where $p_E$-- is the Ejector period) even
for $t_d < 10^8 $ yrs, for weaker fields 
the ``forbidden'' region becomes
wider. Critical period, $p_E$, corresponds to transition from the Propeller
stage to the stage of Ejector, and is about 10-25 seconds for typical
parameters.
The results are dependent on the initial magnetic field, $\mu_0$, 
the ISM density, $n$, and NSs velocity, $V$. So here different ideas can be
investigated. 

In fact the limits obtained above are even stronger than
they could be in nature, i.e. ``forbidden'' regions can be wider, 
because we did not take into account
that NSs can spend 
some significant time (in the case with field decay)
at the propeller stage (the spin-down rate at this stage is very
uncertain, see the list of formulae, for example, in \cite{lp95}
or \cite{l92}). 
The calculations of this effect for
different models of non-exponential field decay were made separately
\cite{pp2001}.

Note that there is another reason due to which
a very fast decay down to small values of $\mu_b$ can also be
excluded, because this would lead to a huge amount of accreting isolated
NSs in drastic contrast with observations. This situation
is similar to the ``turn-off'' of the magnetic field of an INS
(i.e., quenching any magnetospheric effect on the accreting matter). So
for any velocity and density distributions we should expect 
significantly more accreting isolated NSs than we know from ROSAT
observations
(of course, for high velocities X-ray sources will be very dim, but close
NSs can be observed even for velocities $\sim 100$ km s$^{-1}$).

\section*{Log N -- Log S distribution}

In this section  we briefly present our new results on INSs, \cite{p2000b}.

We compute and compare the $\log N$ -- $\log S$ distribution of both
accreting and cooling NSs, to establish the relative
contribution of the two populations to the observed number counts.
Previous studies derived the $\log N$ -- $\log S$
distribution of accretors (\cite{tc91}; \cite{mb94};
\cite{mann96}) assuming  a NSs velocity distribution rich in slow
stars ($v< 100 \ {\rm km\, s}^{-1}$). More recent measurements of pulsar
velocities (e.g. \cite{ll94})
and upper limits on the observed number of accretors in ROSAT surveys 
point, however, to a larger NS mean velocity
(see \cite{t2000} for a critical discussion).
Recently in \cite{nt99} the authors
compared the number count distribution of the
ROSAT isolated NS  candidates  with those of accretors and coolers.
In \cite{p2000b} we address these issues in  greater detail, also
in the light of the latest contributions to the modeling of the evolution
of Galactic NSs.

Our main results for AINSs are presented in Fig.~5 and Fig.~6.

\begin{figure}[h!]
\epsfxsize=10cm
\centerline{\rotate[r]{\epsfbox{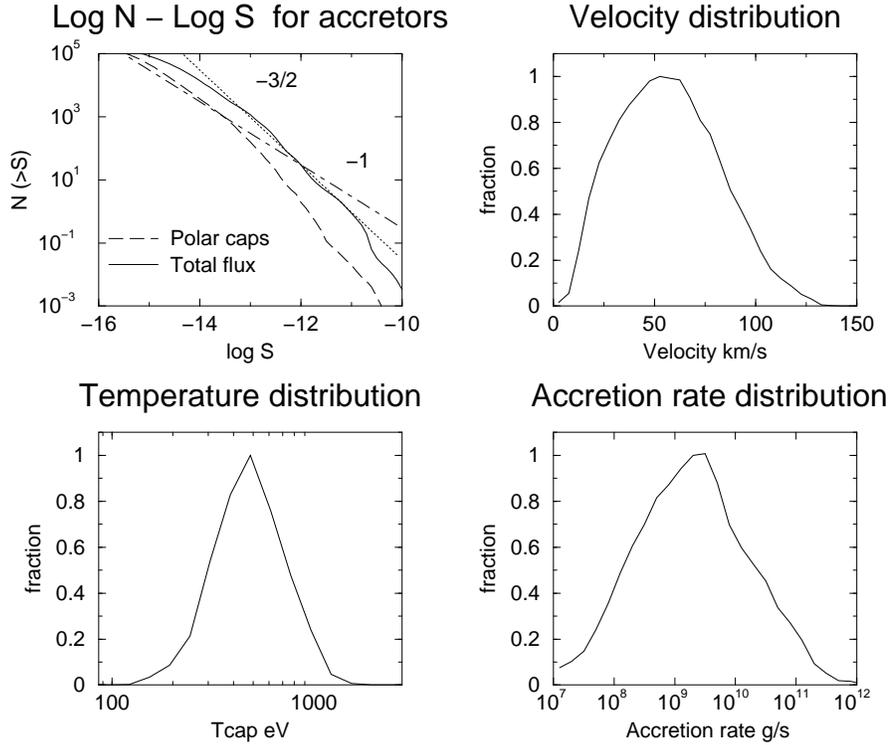}}}
\caption{Upper left panel: the $\log N$ -- $\log S$ distribution for
accretors within 5 kpc from the Sun. The two curves refer to total emission
from the entire star surface and
to polar cap emission in the range 0.5-2 keV; two straight lines with slopes -1 and -3/2
are also shown for comparison. From top right to bottom right: the velocity,
effective temperature and accretion rate distributions of accretors; all
distributions are normalized to their maximum value.}
\end{figure}

In Fig.~5 we show main parameters of accretors in our calculations:
Log N -- Log S distribution, distributions of velocity, accretion rate and 
temperature (for polar cap model).

In Fig.~6 we present joint Log N - Log S for accretors in our calculations,
observed candidates, naive estimate from \cite{p2000a}, and our calculations
for coolers in a simple model of local sources \cite{p2000b}.

Using  ``standard'' assumptions
on the velocity, spin period and magnetic field parameters,
the accretion scenario can not explain the observed properties of the six
ROSAT candidates.

A key result of our  statistical analysis is that
accretors should  eventually become more abundant than coolers
at fluxes below
$10^{-13}$ erg cm$^{-2}$ s$^{-1}$.

\begin{figure}[h!]
\epsfxsize=10cm
\centerline{\rotate[r]{\epsfbox{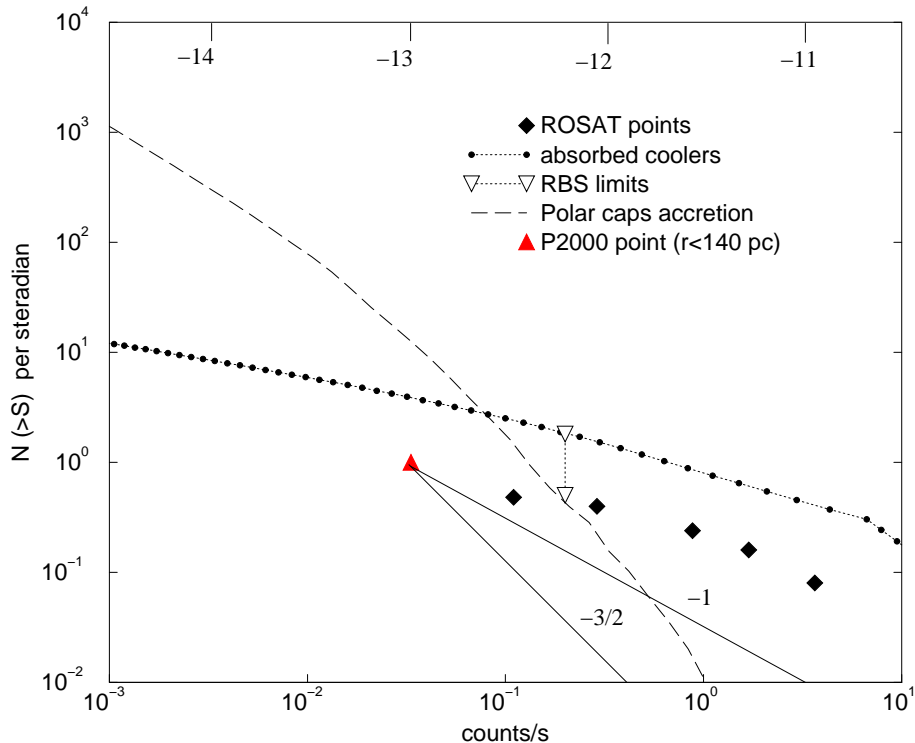}}}
\caption{Comparison 
of the log $N$ -- log $S$ distributions for accretors and coolers together
with observational points, the naive log$N$ -- log$S$ from
P2000a and the ROSAT Bright Survey (RBS) limit.
The scale on the top horizontal axes gives the flux in erg cm$^{-2}$s$^{-1}$.}
\end{figure}

\section*{Conclusions}

 INSs are now really {\it hot} objects. In many cases INSs can show
different effects in the most "pure" form: without influence of 
huge accretion rate, for example.

We tried to show how these objects are related with models
of magnetic field decay, and with recent and future X-ray observations.

 Observed candidates propose ``non-standard'' properties of 
NSs. Future observations with XMM (Newton) and Chandra satellites
can give more important facts. 

 INSs without usual radio pulsar activity (SGRs, AXPs, compact X-ray sources
in SNRs, dim ROSAT candidates, Geminga, dim sources in globular clusters) 
together show ``non-standard'' or better say more complete picture of
NS nature. Future investigations are strongly wanted.

\acknowledgements
I wish to thank  my co-authors Monica Colpi, Vladimir Lipunov,
Mikhail Prokhorov, Aldo Treves and Roberto Turolla.
I also thank Organizers for their kind hospitality.
The work was supported through the grant of Russian Foundation for Basic
Research, Scientific Travel Center and INTAS.

\end{article}

\end{document}